\documentstyle[twoside,psfig,mpla1]{article}

\renewcommand{\thefootnote}{\fnsymbol{footnote}}

\begin{document}

\def    \beq    {\begin{equation}} 
\def    \eeq    {\end{equation}}
\def    \lf     {\left (} 
\def    \rt     {\right )}
\def    \a      {\alpha} 
\def    \r      {\rho}
\def    \th     {\theta}
\def    \lm     {\lambda}
\def    \D      {\Delta}
\def    \Li     {\rm Li_4}
\def    \rg     {\sqrt{-g}}
\def    \pl     {\partial}
\def    \del    {\nabla}
\def    \comma  {\; , \; \;}

\setlength{\textheight}{7.7truein} 

\runninghead{Phases of Thermal Super Yang-Mills}
{Phases of Thermal Super Yang-Mills}

\normalsize\textlineskip
\thispagestyle{empty}
\setcounter{page}{1}

\copyrightheading{}			

\vspace*{0.88truein}

\fpage{1}
\centerline{\bf PHASES OF THERMAL SUPER YANG-MILLS}
\baselineskip=13pt
\vspace*{0.37truein}
\centerline{\footnotesize {MAULIK K. PARIKH
\footnote{m.parikh@phys.uu.nl} }}
\baselineskip=12pt
\centerline{\footnotesize\it 
Spinoza Institute, University of Utrecht}
\baselineskip=10pt
\centerline{\footnotesize\it 
P. O. Box 80 195, 3508 TD Utrecht, The Netherlands}

\vspace*{10pt}


\vspace*{0.21truein}
\abstracts{
We review the thermodynamics of the confined and unconfined phases of
${\cal N} = 4$ super Yang-Mills at large N on a three-sphere,
focussing especially on the confinement-deconfinement transition. We
determine an N-dependent phase boundary and
point out some directions for future work.}{}{}


\setcounter{footnote}{0}
\renewcommand{\thefootnote}{\alph{footnote}}

\vspace*{1pt}\textlineskip	
\vspace*{-0.5pt}

\section{Confinement in Super Yang-Mills Theory}
\noindent
The correspondence\cite{adscft} between type IIB string theory in 
anti-de Sitter space and certain
conformal field theories has allowed us to
study those gauge theories in regimes that were
previously intractable.
In this paper, we review some thermodynamic aspects of this duality
when the conformal field theory is four-dimensional 
${\cal N} = 4$ super Yang-Mills with gauge group SU(N) for large rank N.

The duality is holographic; it relates a CFT on the boundary to bulk
manifolds with the right symmetries. For the ${\cal N} = 4$ SU(N) theory,
there are at least two such geometries: $AdS_5 \times S^5$ and a spacetime
in which the $AdS_5$ is replaced with a Hawking-Page black hole (an
uncharged black hole in an anti-de Sitter background). These
two solutions correspond respectively to the confined and 
unconfined phases in the gauge theory,\cite{thermalphase} as we shall see.
Our aim is to use the correspondence to study the confinement-deconfinement 
transition in the gauge theory.\footnote{This review overlaps
with the paper hep-th/0002031, by the author and D. Berman.\cite{confads}}

The gauge theory action is
\beq 
S = \int d^4 x \rg ~ {\rm Tr}  \lf  -{1 \over 4 g^2}
F^2 + {1 \over 2} \lf D \Phi \rt ^2 + {1 \over 12} R \Phi^2 +
\bar{\psi} \, / \! \! \! \! D \psi \rt \; .	\label{symaction}
\eeq 
All fields are in the
adjoint representation of $SU(N)$. The six scalars, $\Phi$,
transform under $SO(6)$ R-symmetry, while the four Weyl fermions, $\psi$,
transform under $SU(4)$, the spin cover of $SO(6)$. The scalars are
conformally coupled; otherwise, all fields are massless.

This theory is conformal at the quantum level. 
In order to induce interesting phase behavior 
one has to break conformality. We can do this 
by considering the theory at nonzero
temperature, $T$, on a finite volume three-sphere of radius $r_0$. 
Then conformal invariance dictates that thermodynamic properties can depend 
only on the dimensionless quantity $T r_0$ or, in the microcanonical
version, $Ur_0$ where $U$ is the energy of the system.
This implies in particular that the infinite volume limit is equivalent 
to the infinite temperature limit; on $R^3$ the theory is always
in its high temperature phase. Hence to study the 
confined phase we consider the theory on a finite sphere.

However, ``confinement'' on $S^3$ is slightly tricky. One cannot 
envision a single quark on $S^3$ as there is nowhere for the flux to go
(because $\pi_2 (S^3 - \{0\}) \equiv 0$). Instead, we can try to express 
confinement in terms of temporal Wilson loops (Polyakov loops). 
A temporal loop, $C$, traces one end of an open string, whose other end 
terminates on the stack of $N$ D3-branes, that is, in the bulk. So we 
should consider all string worldsheets that extend from a timelike circle on
the boundary into the bulk. The expectation value over these tube-shaped
surfaces serves as an order parameter for the spontaneous breaking of
the $Z_N$ center of $SU(N)$:
\beq
\left < W[C] \right > = {1 \over N} \left < {\rm Tr} ~ P ~ \exp \oint_C \! A 
\right > \, \sim \, e^{- \beta F} \; ,
\eeq
where $P$ denotes path-ordering, $A$ is the gauge field, and the trace is 
in the fundamental representation (to be precise, one should also include
the fermions and scalars). The expectation value is taken with respect to
the action, Eq. (\ref{symaction}). Here we have written it
in terms of the free energy. Loosely speaking, adding a Wilson loop 
is like adding a quark;\cite{juanwilson,soojong} then since a 
zero expectation value costs an infinite amount of free energy, it
corresponds to confinement. Now the topology of the 
black hole is $B^2 \times S^3$, while
the topology of thermal AdS space is $S^1 \times B^4$. Then the argument 
goes that, since $C$ belongs to a nontrivial first homology class for the
thermal AdS spacetime, it cannot be the boundary of a worldsheet. 
Hence the expectation value of the
Wilson loop is zero and so thermal AdS corresponds to the confined phase.
By contrast, for the black hole topology, $C$ could be one boundary 
of a string worldsheet
so in general the Wilson loop has nonzero expectation value.
However, there is a subtlety, first noted by Witten.\cite{thermalphase} IIB
string theory has an NS-NS two-form that, with vanishing field strength,
can be added to the background at no cost in energy. 
Thus the Wilson loop acquires a periodic phase factor,
and summing over all values of the phase gives an expectation value 
of zero for the black hole geometry too. So Wilson loops do not seem to 
characterize the phase of the gauge theory on $S^3$.

To check that the different gravitational backgrounds indeed correspond to
different phases of the gauge theory one could look for other evidence, for 
example by comparing thermodynamic quantities computed on either side of
the correspondence.
Unfortunately, at present it is only possible to do direct calculations in
the gauge theory at weak coupling. Consider then the free field limit
of super Yang-Mills. In the high-energy regime which 
dominates the state counting, the spectrum of free fields on a sphere 
is essentially that of blackbody radiation in flat space, with
$8N^2$ bosonic ($2N^2$ for the gauge bosons, and $6N^2$ for the scalars) 
and $8N^2$ fermionic degrees of freedom. The entropy is therefore
\beq 
S_{\rm CFT} =
{2 \over 3} \pi^2 N^2 \, V_{\rm CFT} \, T^3_{\rm CFT} \; . \label{Scft} 
\eeq
Since the fields have been taken to be noninteracting, this is obviously
the unconfined phase. In the next section, we shall see that anti-de
Sitter black holes have almost exactly the same thermodynamics.

\section{AdS Black Holes}
\noindent
The line element of the dual ``Schwarzschild'' black hole\cite{hawkingpage} 
in $AdS_5 \times S^5$ is
\beq 
 ds^2 = - \lf 1 - {2 M G_5 \over r^2} + r^2 l^2 \rt dt^2 +
\lf 1 - {2 M G_5 \over r^2} + r^2 l^2 \rt^{\! \! -1} dr^2  + r^2 d
\Omega^2_3 + l^{-2} d \Omega^2_5 \; ,  \label{HPds2}
\eeq 
where $G_5$ is the five-dimensional Newton constant, and 
$l$ is the inverse radius of both the five-sphere and AdS, 
so that the five-dimensional cosmological
constant is $\Lambda = -6l^2$. The geometry has a Ricci tensor proportional
to the metric and a nonvanishing Weyl tensor. 

The black hole horizon is at $r= r_+$ where 
\beq  r_+^2 = {1 \over 2 l^2} \lf -1  + \sqrt{1 + 8M G_5
\, l^2} \rt \; . \label{r+}  
\eeq
When $r_+ l \ll 1$, the black hole could become unstable to localization on
the $S^5$ by an analog of the Gregory-Laflamme mechanism.\cite{ruth} 
A necessary condition\cite{gary}
for instability is that the entropy of the localized
black hole be larger than the entropy of a black hole uniformly spread over
the five-sphere. A straightforward computation then shows 
that a localization instability could exist
for very small black holes with $r_+ l \ll 1$.
Here we shall work with stable black holes that have $r_+ l > 1$.

To study the black hole's thermodynamics, we Euclideanize the 
metric. The substitution $\tau = i t$
makes the metric positive definite and, by the usual removal of the
conical singularity at $r_+$, yields an inverse temperature of the black
hole given by the period of $\tau$:
\beq
\beta_{\rm BH} = {2 \pi r_+ \over 1 + 2r_+^2 l^2} \; .\label{invtemp}
\eeq
The entropy is
\beq  
S_{\rm BH} = {A \over 4 G_5} = {\pi^2 r_+^3 \over 2 G_5} \; , \label{Sbh}   
\eeq  
where $A$ is the ``area'' (that is, three-volume) of the horizon.
The mass above the anti-de Sitter background is  
\beq 
U_{\rm BH} 
= {3 \pi \over 4} M = {3 \pi \over 8 G_5} r_+^2 \lf 1 + r_+^2 l^2 \rt  \;
. \label{Ubh} 
\eeq 
This is the AdS equivalent of the ADM mass, or
energy at infinity. (Actually if the black hole is to be considered at thermal
equilibrium it should properly be regarded as being surrounded by a thermal
envelope of Hawking particles. Because of the infinite blueshift at the
horizon, this envelope contributes a formally infinite energy. 
Here we shall neglect this infinite energy as unphysical, absorbed perhaps
by a renormalization of the Newton constant. But there could be finite
regularization-scheme-dependent corrections of order zero in $G_5$.) 

We can now also write down the free energy:
\beq
F_{\rm BH} = {\pi r_+^2 \over 8 G_5} \lf 1 - r_+^2 l^2 \rt \; . \label{Fbh}
\eeq
Eqs. (\ref{invtemp}-\ref{Fbh}) satisfy the first law of thermodynamics.
It is interesting to note that, by formally dividing both the free energy
and the mass by an arbitrary volume, one obtains an equation of state: 
\beq 
p = {1 \over 3} {r_+^2 l^2 - 1 \over r_+^2 l^2 + 1} \, \rho \; ,
\eeq 
where $p = - F/V$ is the pressure, and $\rho$ is the energy density. 
In the limit $r_+ l \gg 1$ this equation becomes 
\beq  
p = {1 \over 3} \rho  \; , 
\eeq 
as is appropriate for the equation of state of a conformal
theory. This suggests that if a conformal field theory is to
reproduce the thermodynamic properties of this gravitational solution,
it has to be in the $r_+ l \gg 1$ limit.

To express the CFT parameters $N$, $T_{\rm CFT}$, and $V_{\rm CFT}$ 
in terms of black hole quantities, we implement holography
by taking the physical data for the CFT from the boundary of the black hole
spacetime. At fixed $r \equiv r_0 \gg r_+$, the boundary line element
tends to  
\beq  
ds^2 \to r_0^2 \left [ - l^2 dt^2 + d \Omega_3^2 \right ] \; ,
\eeq 
which has a spatial volume of
\beq  V_{\rm CFT} = 2 \pi^2
r_0^3 \; .   
\eeq 
The conformal field theory temperature is the physical temperature 
at the boundary:
\beq
T_{\rm CFT} = {T_{\rm BH} \over \sqrt{-g_{tt}}} \approx 
{1 \over l r_0} {1 + 2 r_+^2 l^2 \over 2 \pi r_+} \; . \label{Tcft}
\eeq
To obtain an expression for $N$, we invoke the
AdS/CFT correspondence. This 
relates $N$ to the radius of $S^5$ and the cosmological
constant: 
\beq  R^2_{S^5} = \sqrt{4 \pi g_s {\a '}^2 N} = {1 \over
l^2} \; .  
\eeq 
Then, since
\beq
(2 \pi)^7 g_s^2 {\a '}^4 = 16 \pi G_{10} =  16 {\pi^4 \over l^5} G_5 \; , 
\eeq 
we have 
\beq  
N^2 = {\pi\over 2 l^3 G_5} \; . \label{N} 
\eeq
With these substitutions, we see that the CFT entropy in the unconfined
phase, Eq. (\ref{Scft}), is (almost) the same as the black hole entropy,
Eq. (\ref{Sbh}), in the limit $r_+ l \gg 1$:
\beq
S_{\rm CFT} \sim  {4 \over 3} S_{\rm BH} \; . \label{S}
\eeq
Similarly, the red-shifted energy of the conformal field theory matches the
black hole mass, modulo a coefficient:
\beq 
U^{\infty}_{\rm CFT} = \sqrt{-g_{tt}} 
{\pi^2 \over 2} N^2 \, V_{\rm CFT} \, T^4_{\rm CFT}
\sim {\pi \over 2} r_+^4 l^2 \sim {4 \over 3} U_{\rm BH} \; , \label{U}
\eeq 
where $U^{\infty}_{\rm CFT}$ is the conformal field theory energy
red-shifted to infinity, and we have again taken the $r_+ l \gg 1$ limit.
At this level, the correspondence only goes
through in this high temperature limit. Since the only two scales in the 
thermal conformal field theory are $r_0$ and $T_{\rm CFT}$, high temperature 
means that $T_{\rm CFT} \gg 1 / r_0$, allowing us to neglect 
finite-size effects.

The mysterious $4/3$ discrepancy in Eqs. (\ref{S}) and (\ref{U}) 
is usually construed to be
an artifact of having calculated the gauge theory entropy in the free
field limit rather than in the limit of strong 't Hooft coupling, $\lm \gg 1$,
as required by the correspondence; 
intuitively, one expects the free energy to decrease
when the coupling increases. The $4/3$ factor was first noticed in the context 
of D3-brane thermodynamics.\cite{fourthirds} The thermodynamic
matching has been extended 
to rotating black holes\cite{5dadskerr,holorotbh,harvey} and
their field theory duals, Yang-Mills with angular momentum for which,
interestingly, the tantalizing $4/3$ factor remains unchanged.\cite{holorotbh}

For our present purposes, the most interesting aspect is the 
N-dependence of the result. The dependence on $N^2$ indicates that the 
conformal field theory is in its unconfined phase; the $N^2$ species of 
free gluons make independent contributions to the free energy. We shall
see in the next section that the thermodynamics of the confined phase 
is rather different.

\section{A Hot Bath in AdS}
\noindent
Now consider a gas of thermal radiation in anti-de Sitter space. This is
dual to the confined phase of super Yang-Mills. (Curiously,
attempts to formulate confinement in terms of anti-de Sitter space date
back at least to the 1970's.\cite{salam}) The thermal gas is necessary
to ensure that the energy measured with respect to the AdS background is
nonzero. The energy eigenstates of $AdS_5$ are:\cite{isham}
\beq
\Psi_{\omega j m n} (r, t, \theta, \phi, \psi) = 
N_{\omega j} \, \exp \lf -i \, \omega \, l \, t \rt \,
\sin ^j \rho \,
C^{j+1}_{\omega -j-1}(\cos\rho) \, Y^{m n }_j (\theta, \phi, \psi) \; ,
\eeq
where $C^{p}_{q} \, (x)$
are Gegenbauer polynomials, $Y^{m n}_j (\theta, \phi, \psi)$ 
are the spherical harmonics in five-dimensional
spacetime (with total angular momentum number $j$), and 
$\rho \equiv \arctan (rl)$. Here $\omega$ is an integer 
satisfying the condition $\omega - 1 \geq j \geq |m|,|n|$.  
Hence the spectrum is quantized in units of $l$, 
the inverse radius of AdS. 
Since this is also the quantum of excitations of the five-sphere, 
we should consider thermodynamics over the full ten-dimensional space.
The appropriate line element is therefore
\beq 
ds^2 = - \lf 1 + r^2 l^2 \rt dt^2 + \lf 1 + r^2 l^2 \rt^{-1} dr^2 + r^2 d
\Omega_3^2 + l^{-2} d \Omega_5^2 \; . \label{AdS}
\eeq 
To obtain a thermal field theory, we again Euclideanize the metric. The
periodicity of $\tau = it$ is then the inverse 
(asymptotic) temperature, $T^{-1}_{\rm AdS}$, of the theory; the absence of
a horizon means that $T_{\rm AdS}$ is an arbitrary parameter. However, the
relevant temperature for thermodynamics in the bulk is not $T_{\rm AdS}$, but
the local, redshifted, temperature:
\beq 
T_{\rm local} = {T_{\rm AdS} \over \sqrt{-g_{tt}}} 
= {T_{\rm AdS} \over \sqrt{ 1+ r^2 l^2}} \; . \label{redshift}
\eeq
To calculate thermodynamic quantities we foliate spacetime into (timelike)
slices of constant local temperature. Extensive thermodynamic quantities
are then computed by adding the contribution of each such hypersurface.
 
The local energy density of the thermal gas of radiation is
\beq 
\rho_{\rm local} = \sigma T^{10}_{\rm local} \; , \label{rho}
\eeq 
where we have neglected infrared effects due to curvature or nonconformality.
(This assumes a thermodynamic limit; an exact statistical mechanical
computation would be preferable.) Here $\sigma$ is the 
ten-dimensional supersymmetric 
generalization of the Stefan-Boltzmann constant, which is approximated
by its flat space value:
\beq 
\sigma = {62 \over 105} \pi^5 \; ,
\eeq
where we have included a factor of 128, the number of 
massless bosonic physical degrees of freedom of IIB supergravity.

The total ``ADM'' energy-at-infinity of a gas contained in a ball
of radius $r_0$ is then
\beq
U_{\rm gas}^{\infty} 
= \sigma {2 \pi^5 \over l^5} \int_0^{r_0} T^{10}_{\rm local} 
\sqrt{-g_{tt}} \sqrt{g_{rr}} \,
r^3 dr \equiv \sigma V_{\rm eff} (r_0) T_{\rm AdS}^{10} \; . 
\label{Ugas} 
\eeq
Here the additional blueshift factor of $\sqrt{-g_{tt}}$
converts the local (fiducial) energy into an ADM-type 
energy, comparable to Eq. (\ref{Ubh}). We have also defined an 
effective volume,
\beq 
V_{\rm eff}(r_0)= {2 \pi^5 \over l^9} \lf {2 \over 3} -{ {2+ 3
(r_0 l)^2 } \over {3 \lf 1 + (r_0 l)^2 \rt ^{3/2} }} \rt \; ,
\eeq
which, as $r_0 \rightarrow \infty$, approaches
\beq 
{4 \pi^5 \over {3l^9}} \; .
\eeq
Thermodynamically, anti-de Sitter space behaves as if it had a
finite volume.

Similarly, the other thermodynamic quantities of the thermal bath are
\beq 
F = - {\sigma \over 9} V_{\rm eff} T_{\rm AdS}^{10} \comma 
S = {10 \over 9} \sigma V_{\rm eff} T_{\rm AdS}^9 \; , \label{gasent}
\eeq
consistent with the first law of thermodynamics.
The absence of a $G_5$  in the free energy
indicates, from the CFT point of view, 
that the free energy is of order $N^0$. This is the confined
phase of the theory -- the free energy is of order $N^0$ because the $N^2$
species of gluons have condensed into hadronic color singlets. 

Rewritten in CFT quantities, Eq. (\ref{gasent}) implies that 
$S_{\rm CFT} \sim r_0^9 \, T^9_{\rm CFT}$ 
since $T_{\rm CFT} \approx T_{\rm AdS} / (l r_0)$.
The nine-dimensional volume is somewhat puzzling for
a three-dimensional gauge theory. It reflects the fact that the 
QCD (SYM) string is really a type IIB string which naturally lives in nine 
spatial dimensions. It has been suggested that
the extra dimensions in which the open string worldsheet 
bounded by a Wilson loop can extend are akin to 
Liouville dimensions.\cite{adscft,juanwilson,cave,nadavgross}

\section{The Crossover}
\noindent
In the microcanonical approach that we shall now pursue, the
contributions to the partition function come from both the black
hole and the gas in AdS where the energy of the 
gas and black hole are taken to be the same.
Which of these two thermodynamic phases the system is found in is
determined, in the saddle point approximation, by the relative values of
the respective Euclidean classical actions:
\beq
Z(U) = e^{-I_{\rm BH}(U)} + e^{-I_{\rm gas}(U)} \; . \label{part}
\eeq
The action of the black hole is
simply the Einstein-Hilbert action with a negative cosmological constant.
This is proportional to the proper volume of spacetime and so
needs to be regulated. 
We will refer to Euclideanized anti-de Sitter space as thermally-identified
AdS, to distinguish it from thermal AdS with a gas of radiation.
Thermally-identified AdS was chosen as the zero of energy in the black hole
mass formula, Eq. (\ref{Ubh}). We choose it also as the zero of action.
A finite black hole action is obtained by subtracting the
(also infinite) action for thermally-identified anti-de Sitter 
space in which the
hypersurface at a constant large radius has the same intrinsic 
geometry as a hypersurface at the same radius in the black hole 
background.\cite{thermalphase,hawkingpage} 
The resultant regularized black hole action is
\beq  
I_{\rm BH} = -{1 \over 16 \pi G_5} \int d^5 x \sqrt{-g} \lf R + 12 l^2 \rt
= {\pi^2 r_+^3 \over 4 G_5} {1 - r_+^2 l^2 \over 1 + 2 r_+^2 l^2} \; . 
\label{IBH}
\eeq
The comparable value of the action for 
the thermal gas in AdS is just the action of the gas itself, namely
$F/T$:
\beq
I_{\rm gas} = -{1 \over 9} \sigma V_{\rm eff} T_{\rm AdS}^{9} \; . \label{Igas}
\eeq

The qualitative thermodynamic behavior of the system is determined by
the phase whose action dominates the partition function Eq. (\ref{part}).
The system can change phase by tunneling. 
This is not a phase transition in the Landau sense of a singularity in the
free energy; in general, such mathematically strict 
phase transitions are not possible in finite volume because 
the partition function is then a finite sum 
which is therefore analytic in all couplings.
Rather, the changes in phase are smooth crossovers that 
occur with a probability
\beq
\Gamma \sim \exp (\Delta I) = \exp (- \, {\cal O} \lf N^2 \rt ) \; ,
\eeq
where $\Delta I$ is the difference in the Euclidean action between the two
phases. As $N$ goes to infinity the crossover does not occur. At the
crossover between the two phases, the action for the
gas and the black hole are the same. Moreover, energy must be conserved.
Thus at the crossover, we have
\beq
U^{\rm local}_{\rm gas} = U^{\rm local}_{\rm BH} \comma
I_{\rm gas} = I_{\rm BH} \; .
\eeq
Note that, since the two phases cannot be in physical contact, 
the physical temperature is not required to be the same for the two
phases.

Solving these equations yields $N^2$ as a function of the dimensionless
quantity $x \equiv r_+ l$ at the crossover:
\beq
N^2 = {31 \over 2^5 \cdot 3^{13} \cdot 5 \cdot 7} \, 
{(1+x^2)^9 \cdot (1+2x^2)^{10} \over (x^2 - 1)^{10} \cdot x^{12}} \; . 
\label{cross}
\eeq
This is the equation we want.\cite{confads} 
For given $N$, it determines the value
of $x$ at which the phase changes. Using Eq. (\ref{Tcft}), we see that
$x$ is related to the dimensionless quantity $T_{\rm CFT} \, r_0$ (or better,
to $U_{\rm CFT} \, r_0$) as predicted from conformal invariance. 
Before discussing
the crossover curve, we comment on some of the underlying approximations.

In using Eq. (\ref{AdS}), 
we have omitted the back-reaction of the gas on the
metric. Back-reaction can reliably be
neglected when the matter term in the parenthesis is smaller than the
cosmological term. For an energy density given by Eq. (\ref{rho}) and
a total energy matched to that of the black hole phase, Eq. (\ref{Ubh}),
the condition $G_5 \rho < |\Lambda|$ in the $t-t$ Einstein equation
amounts to
\beq
{9 \pi \over 32} \, x^2 (1 + x^2) \, l^2 < 6 \, l^2 \; ,
\eeq
and we see that the matter term becomes dominant at large $x$, and is not
entirely negligible even near $x = 1$. At high temperature, therefore,
Eq. (\ref{cross}) becomes unreliable.
The exact form of Eq. (\ref{cross}) would also be modified by 
including the correct Stefan-Boltzmann constant for anti-de Sitter space.
More importantly, since the temperature can be of the order of the 
ground state energy, we need a statistical mechanical derivation of the
energy density. And finally, when $N$ is small,
the supergravity approximation itself breaks down.

\begin{figure}[ht]
\vspace*{13pt}
\centerline{\psfig{file=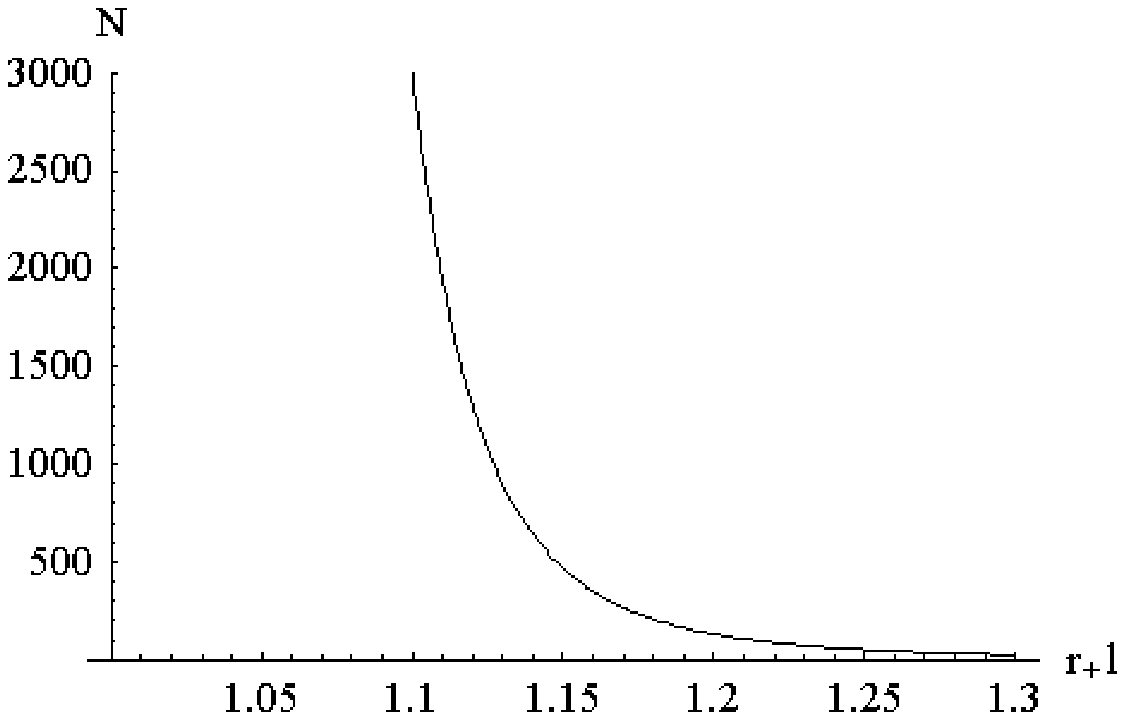,width=100mm}}
\vspace*{13pt}
\fcaption{$N$ vs. $r_+ l$ at the crossover.}
\end{figure}

Despite these caveats, Eq. (\ref{cross}) seems to capture the correct
qualitative behavior. In Fig. 1, we plot $N$ near the crossover for 
$x \sim 1$. 
The region below the crossover curve is dominated by the confined 
or AdS gas phase, whereas the region above is dominated by 
the unconfined or black hole phase. As the temperature increases, the
graph confirms our expectation that the gauge theory recovers conformality.
As $N$ goes to infinity, we recover the result of Witten,\cite{thermalphase}
that the transition occurs at $r_+ l = 1$. We see that in fact this is
a very good approximation for finite (but large) N.

\section{Conclusion and Outlook}
\noindent
We have determined the N-dependent curve that separates the confined
and unconfined phases of thermal super Yang-Mills theory on a three-sphere,
by making use of the AdS/CFT correspondence. A direct derivation of 
such a phase diagram would have been difficult.

Plenty of work remains to be done. There are several 
technical assumptions that need to be tightened: 
the effect of the backreaction of the gas on 
the metric (if not exactly, then at linear order in $G_5$), a statistical
mechanical (as opposed to thermodynamic) expression for the energy of the
gas in AdS, the AdS value of the Stefan-Boltzmann constant, finite
one-loop corrections (perhaps using zeta-function regularization) 
to the black hole background, etc. These technical improvements are
important because some of the approximations made here are vulnerable
precisely in the regime of greatest interest. The inclusion of other parameters
such as rotation or R-charge\cite{steve} 
are obvious generalizations; these more 
complicated solutions might have phase diagrams with interesting properties.
One could also study what happens with different boundaries, or in different
dimensions. For $AdS_3$, much is known about the corresponding conformal 
theory; perhaps, this could serve as a test.

The most interesting -- if not most promising -- direction is to ask whether
there are any implications for real multicolor (large N) QCD. 
Super Yang-Mills differs from ordinary QCD in several ways: it is
supersymmetric, of course, also conformal and has extra 
spin zero and spin one-half adjoint fields. One possible approach to QCD is 
as follows.\cite{thermalphase}
Consider the six-dimensional conformal theory with $(0,2)$ supersymmetry
that is dual to $AdS_7 \times S^4$ (which is the near-horizon geometry of 
N parallel coincident M5-branes).
Now compactify on an $S^1$ with supersymmetry-breaking anti-periodic
boundary conditions for the fermions;
since the fermions have no zero modes, they acquire a mass at tree level
proportional to the inverse radius of the circle. The mass of the scalars
is not protected by any symmetry, so in general we expect them to acquire
a mass at one-loop. Compactifying on a second circle then gives 
ordinary large N pure glue QCD at zero temperature. 
Unfortunately, the glueballs in this
theory typically have an excitation energy of the same order as the
Kaluza-Klein modes, suggesting that real QCD may remain out of reach.

It would be interesting to see if our analysis could be pursued in some of
these directions.

\nonumsection{Acknowledgment}
\noindent
The author is supported by the 
Netherlands Organization for Scientific Research (NWO).

\nonumsection{References}

\end{document}